\documentclass[prb,twocolumn,showpacs,preprintnumbers,amsmath,amssymb,citeautoscript]{revtex4-1}
\usepackage{graphicx}
\usepackage{dcolumn}
\usepackage{bm}
\usepackage{color}

\begin{document}

\title{Disorder-induced topological change of the superconducting gap structure in iron pnictides}

\author{Y.~Mizukami$^{1,2}$}
\author{M.~Konczykowski$^{3}$}
\author{Y. Kawamoto$^1$}
\author{S. Kurata$^{1,2}$}
\author{S.~Kasahara$^1$}
\author{K.~Hashimoto$^{1,4}$}
\author{V.~Mishra$^5$}
\author{A.~Kreisel$^6$}
\author{Y.~Wang$^6$}
\author{P.\,J.~Hirschfeld$^6$}
\author{Y.~Matsuda$^1$}
\author{T.~Shibauchi$^{1,2}$}

\affiliation{
$^1$Department of Physics, Kyoto University, Sakyo-ku, Kyoto 606-8502, Japan\\
$^2$Department of Advanced Materials Science, University of Tokyo, Kashiwa, Chiba 277-8561, Japan\\
$^3$Laboratoire des Solides Irradi\'es, CNRS-UMR 7642 \& CEA-DSM-IRAMIS, Ecole Polytechnique, F 91128 Palaiseau cedex, France\\
$^4$Institute for Materials Research, Tohoku University, Aoba-ku, Sendai 980-8577, Japan\\
$^5$Materials Science Division, Argonne National Laboratory, Lemont, IL-60439, USA\\
$^6$Department of Physics, University of Florida, Gainesville, FL 32611, USA
}

\date{\today}


\begin{abstract}
{\bf
In superconductors with unconventional pairing mechanisms, the energy gap in the excitation spectrum often has nodes, which allow quasiparticle excitations at low energies. In many cases, e.g. $d$-wave cuprate superconductors, the position and topology of nodes are imposed by the symmetry, and thus the presence of gapless excitations is protected against disorder. Here we report on the observation of distinct changes in the gap structure of iron-pnictide superconductors with increasing impurity scattering. By the successive introduction of nonmagnetic point defects into BaFe$_2$(As$_{1-x}$P$_x$)$_2$ crystals via electron irradiation, we find from the low-temperature penetration depth measurements that the nodal state changes to a nodeless state with fully gapped excitations. Moreover, under further irradiation the gapped state evolves into another gapless state, providing bulk evidence of unconventional sign-changing $s$-wave superconductivity. This demonstrates that the topology of the superconducting gap can be controlled by disorder, which is a strikingly unique feature of iron pnictides.
}
\end{abstract}

\maketitle

When  repulsive electron-electron interactions are strong, a sign change in the superconducting order parameter (or the gap function) often leads to some energy gain for  electron pairing \cite{Scalapino95,Hirschfeld11}. The positions of the gap nodes in momentum ${\bf k}$ space, at which the order parameter changes  sign, are determined by the superconducting pairing interactions. In the high transition temperature ($T_{\rm c}$) cuprates, it has been established that the lines of nodes are located along $k_x=\pm k_y$, as expected in the $d_{x^2-y^2}$ symmetry of the order parameter \cite{Tsuei00}. In iron pnictides, on the other hand, the multiband Fermi surface structure leads to several candidates for the superconducting symmetry, including sign-changing $s_\pm$ with and without nodes, sign-preserving $s_{++}$, $d$-wave and time-reversal symmetry breaking $s+{\rm i}d$ \cite{Hirschfeld11,Mazin08,Kuroki09,Graser09,Kontani10,Thomale11}. Different symmetries and details of higher order momentum dependence of the gap yield topologically different nodal structures, such as vertical or horizontal lines of nodes, nodal loops or no nodes.

The effects of nonmagnetic impurities on unconventional superconductivity has been one of the central issues in condensed matter physics \cite{Balatsky06,Alloul09}. For the conventional BCS superconductors, it has been established that the effect is essentially null: the transition temperature is robust against nonmagnetic impurity scattering (the so-called Anderson theorem) and so is the superconducting gap. For the unconventional superconductors, the important aspect of the impurity effects is to mix gaps on different parts of the Fermi surface and thereby smear out the momentum dependence \cite{Mishra09}. In the case of superconducting gap with symmetry protected nodes such as $d$-wave, this averaging mechanism leads to the suppression of the gap amplitude, which enhances the low-lying quasiparticle excitations near the nodal positions. In addition to this, the sign change in the order parameter gives rise to  impurity-induced Andreev bound states, which lead to additional quasiparticle excitations \cite{Hirschfeld93}. Such pair-breaking effects of impurities have been observed, e.g. in Zn-doped YBa$_2$Cu$_3$O$_7$ in the bulk measurements of magnetic penetration depth, where the $T$-linear temperature dependence in the clean-limit $d$-wave superconductivity gradually changes to a $T^2$ dependence at low temperatures with increasing Zn concentrations \cite{Bonn93}.

In sharp contrast, when the nodal positions are not symmetry protected, as in the nodal $s$-wave case, the averaging mechanism of impurity scattering can displace the nodes, and at a certain critical impurity concentration the nodes may be lifted if intraband scattering dominates \cite{Mishra09}, eliminating the low-energy quasiparticle excitations. In the fully gapped state after the node lifting, we have two cases in the multiband superconductors. If the signs of the order parameter on different bands are opposite, residual interband scattering  can give rise to midgap Andreev bound states localized at nonmagnetic impurities that can contribute to the low-energy excitations, provided that the concentration of impurities is enough to create such states. If there is no  sign change,  Anderson's theorem will be recovered, no Andreev states will be created,  and thus no significant further change is expected. Indeed such a difference between nodal sign-changing $s_\pm$ and sign-preserving $s_{++}$ cases has been theoretically suggested by the recent calculations for multiband superconductivity considering the band structure of iron pnictides \cite{Wang13}. Therefore, the impurity effects on the gap nodes and low-energy excitations can be used as a powerful probe for the pairing symmetry of superconductors. 

\begin{figure}[t]
\includegraphics[width=1.0\linewidth]{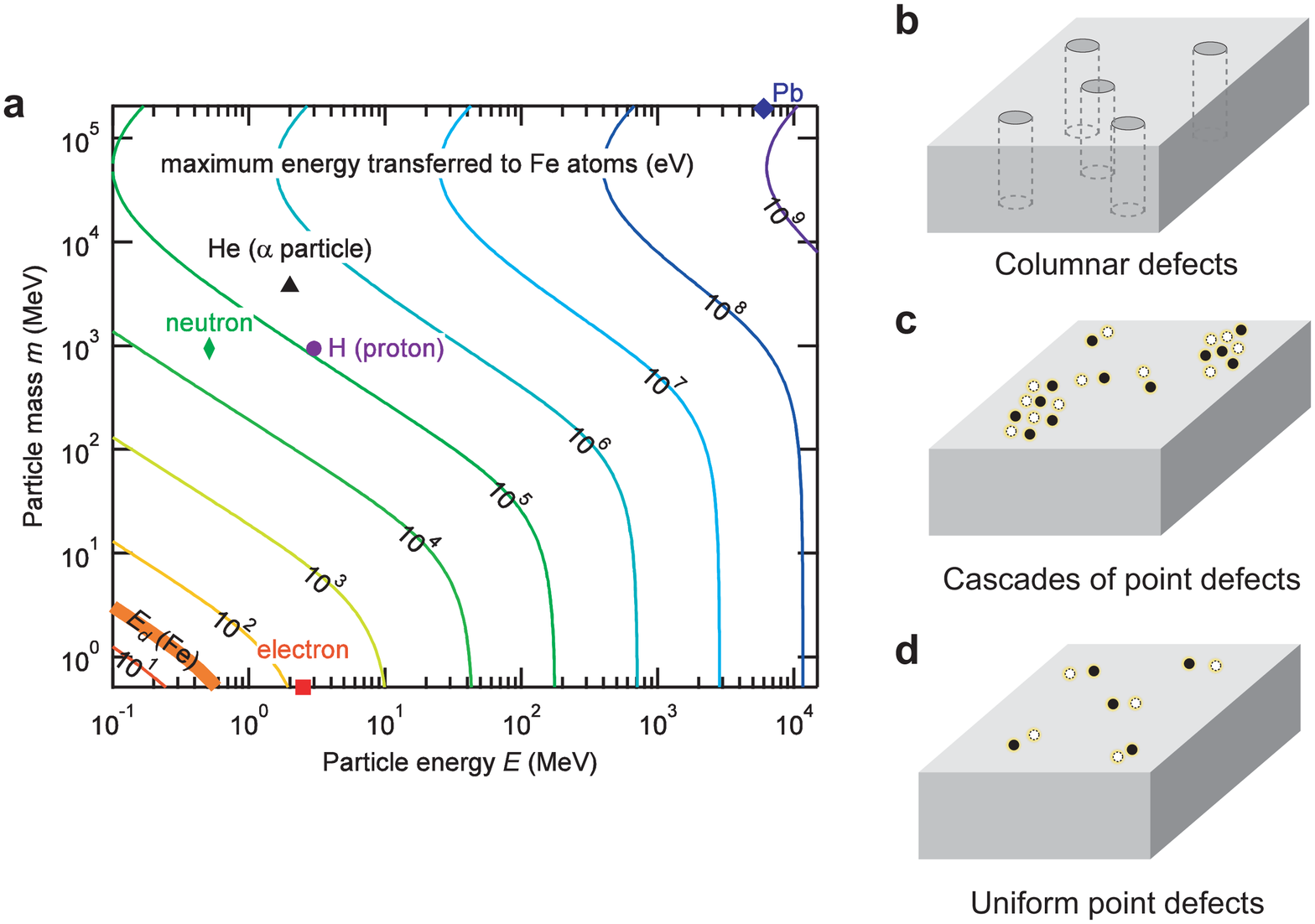}
\caption{{\bf Particle irradiation and created defects.}
{\bf a}, Contour plot of maximum energy transferred to Fe atoms (recoil energy) for irradiated particles with the rest mass $m$ and incident energy $E$. Typical threshold energy $E_d$ for the displacement of Fe atoms from the lattice is marked by thick orange line. 2.5\,MeV electron irradiation used in this study (red square) has orders of magnitude smaller recoil energy than other particle irradiation owing to the small mass. Typical energies for neutron (green diamond), proton (purple circle), $\alpha$ particle (black triangle) and heavy-ion Pb (blue diamond) irradiation are indicated. {\bf b-d}, Schematic illustrations for different types of defects created by particle irradiation. Columnar defects can be created by heavy-ion irradiation ({\bf b}). Particle irradiation with relatively large recoil energies tends to have cascades of point defects due to successive collisions of atoms ({\bf c}). Electron irradiation with a small recoil energy is the most reliable way to obtain uniform point defects ({\bf d}).
}
\label{irradiation}
\end{figure}

In this study, we focus on isovalently substituted system BaFe$_2$(As$_{1-x}$P$_x$)$_2$ close to the optimum composition with $T_{\rm c}\sim30$\,K \cite{Kasahara10}. This system is particularly suitable for the study of the impurity effect on gap structure, because several experiments have indicated that the pristine crystals are very clean and exhibit nodes in the superconducting gap \cite{Shibauchi14}. 
To detect changes in the low-energy quasiparticle excitations, we measure the magnetic penetration depth $\lambda$ at low temperatures,  a fundamental property of superconductors whose $T$ dependence is directly related to the excited quasiparticles. The tunnel diode oscillators (TDOs) in $^3$He and dilution refrigerators operating at 13\,MHz are used to measure the temperature dependence of $\lambda$ down to 0.4\,K and 80\,mK, respectively \cite{Hashimoto10,Hashimoto12}. To introduce the impurity scattering in a controllable way, we employ electron irradiation with the incident energy of 2.5\,MeV, for which energy transfer from impinging electron to the lattice is above the  threshold energy for the formation of vacancy-interstitial (Frenkel) pairs that act as point defects. The long attenuation length and the small recoil energy due to the small mass of electrons is important to create uniformly distributed point defects over the entire crystal with the width of $\sim 30\,\mu$m (Figs.\:\ref{irradiation}a-d). If the recoil energy is too large, one may expect creation of complex defects such as clusters and cascades of point defects and, in the extreme case, columnar tracks, which has been realized by heavy-ion irradiation \cite{Nakajima09} (Figs.\:\ref{irradiation}b and c). Another advantage of electron irradiation is that, unlike chemical substitutions, the defects can be introduced without changing lattice constants, which is quite important as the gap structure may be sensitive to the lattice parameters in iron-based superconductors \cite{Hirschfeld11}.

\begin{figure}[t]
\includegraphics[width=0.75\linewidth]{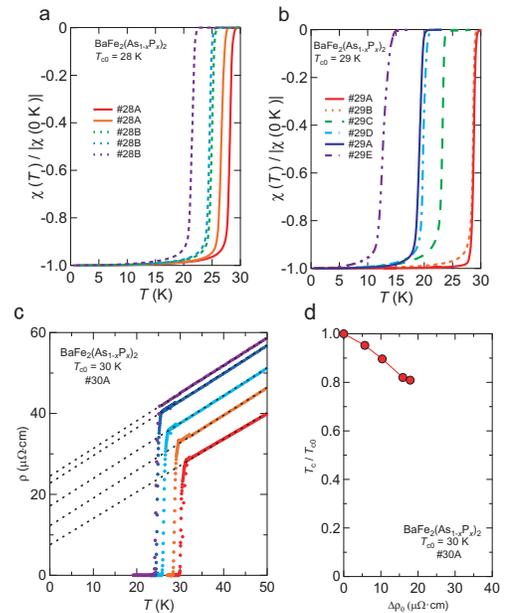}
\caption{{\bf Effect of electron irradiation on the superconducting transition in BaFe$_2$(As$_{1-x}$P$_x$)$_2$ single crystals.}
{\bf a}, Temperature dependence of ac susceptibility at 13\,MHz for crystals (\#28A and \#28B) from a batch of $T_{\rm c0}=28$\,K with irradiated doses of 0, 1.0, 2.0, 4.0, 6.0 C/cm$^2$ with decreasing $T_{\rm c}$. 
{\bf b}, Similar plot for crystals (\#29A, \#29B, \#29C, \#29D, and \#29E) from a batch of $T_{\rm c0}=29$\,K with irradiated doses of 0, 1.5, 2.7, 4.7, 4.9, 8.3\,C/cm$^2$ with decreasing $T_{\rm c}$.
{\bf c}, Temperature dependence of resistivity for a crystal (\#30A) with $T_{\rm c0}=30$\,K with irradiated doses of 0, 1.1, 2.3, 3.8, 4.5\,C/cm$^2$ with decreasing $T_{\rm c}$. Dotted lines are linear extrapolations to zero temperature to estimate changes in residual resistivity $\Delta\rho_0$.
{\bf d}, Transition temperature $T_{\rm c}$ normalized by the pristine value $T_{\rm c0}=30$\,K as a function of $\Delta\rho_0$, determined by the resistivity measurements in crystal \#30A.
}
\label{Tc_dose}
\end{figure}

By successive electron irradiation into clean BaFe$_2$(As$_{1-x}$P$_x$)$_2$  ($x=0.33-0.36$) single crystals with the initial transition temperatures $T_{c0}$ of 28, 29, and 30\,K, we observe a systematic downward shift of $T_{\rm c}$ with increasing defect dose (Figs.\:\ref{Tc_dose}a-d). The transition width in the ac susceptibility data measured by the TDO frequency change (Figs.\:\ref{Tc_dose}a and b) remains almost unchanged after irradiation, which implies a good homogeneity of the  point defects introduced. The temperature dependence of in-plane resistivity $\rho(T)$ measured by the van der Pauw configuration (Fig.\:\ref{Tc_dose}c) shows parallel shifts of $\rho(T)$ after each irradiation in a single crystal. The parallel shifts imply that point defects increase impurity (elastic) scattering with little changes of carrier concentrations and inelastic scattering, which is closely related to the electron correlations. 

We estimate the change in the residual resistivity $\Delta\rho_0$ by extrapolating the normal state data linearly to the zero temperature. The $T_{\rm c}$ reduction rate with respect to $\Delta\rho_0$ is about $-0.3$\,K$\mu\Omega^{-1}$cm$^{-1}$ (Fig.\:\ref{Tc_dose}d), which is comparable to the similar electron irradiation measurements in Ru-substituted BaFe$_2$As$_2$ \cite{Prozorov14}. Previous studies of the $T_{\rm c}$ reduction rate in iron-pnictide superconductors by chemical substitutions \cite{Sato10,Li12,Kirshenbaum12} and by particle irradiation \cite{Prozorov14,Tarantini10,Nakajima10,Taen13} focus on the comparisons with theoretical calculations for $s_\pm$ and $s_{++}$-wave superconductivity \cite{Onari09,Wang13}, but they report various values of the suppression rate. Here we instead focus on changes of low-energy excitations induced by disorder from penetration depth measurements.

\begin{figure}[t]
\includegraphics[width=1.0\linewidth]{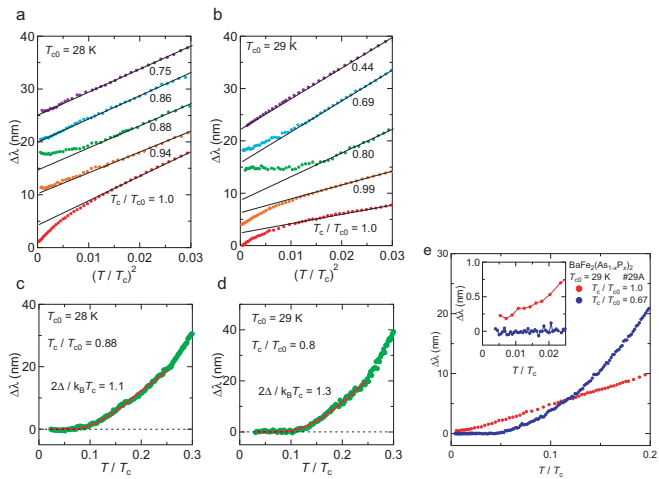}
\caption{{\bf Effect of electron irradiation on the low-temperature penetration depth in BaFe$_2$(As$_{1-x}$P$_x$)$_2$ single crystals.}
{\bf a,b}, Change in the magnetic penetration depth $\Delta\lambda$ plotted against $(T/T_{\rm c})^2$ for the same $T_{\rm c0}=28$\,K samples as Fig.\:\ref{Tc_dose}a ({\bf a}) and for the $T_{\rm c0}=29$\,K samples as Fig.\:\ref{Tc_dose}a ({\bf b}). The same colours are used as Figs.\:2a and b for the corresponding irradiated doses.  Each curve is shifted vertically for clarity. Lines are the $T^2$ dependence fits at high temperatures.
{\bf c,d}, Temperature dependence of $\Delta\lambda$ for
sample \#28B with 2.0\,C/cm$^2$ ({\bf c}) and for sample \#29C with 2.7\,C/cm$^2$ ({\bf d}). Red lines are the fits to the exponential dependence expected in fully gapped superconductors.
{\bf e}, Temperature dependence of $\Delta\lambda$ for
sample \#29A before (red) and after irradiation (blue) of 4.9\,C/cm$^2$ dose. Inset is an expanded view at the lowest temperatures.
 }
\label{lambda}
\end{figure}

In pristine crystals of BaFe$_2$(As$_{1-x}$P$_x$)$_2$, the penetration depth shows a strong temperature dependence at low temperatures (Figs.\:\ref{lambda}a and b), as reported previously \cite{Hashimoto10,Hashimoto12}. The temperature dependence of the change in the penetration depth $\Delta\lambda(T)=\lambda(T)-\lambda(0)$ can be fitted to a power law $T^n$ with the exponent $n<1.5$, indicating that this system has lines nodes in the energy gap. After irradiation, we first find that $\Delta\lambda(T)$ at the lowest temperatures becomes more gentle, or the exponent $n$ increases, and in $T_{\rm c0}=28$\,K crystal we have almost $T^2$ dependence (Fig.\:\ref{lambda}a). Further increase of the defect density results in a flat temperature dependence below $T/T_{\rm c}\sim0.06-0.1$, indicating that the system is changed to a fully gapped state (Figs.\:\ref{lambda}c and d). A fitting to the exponential dependence gives the gap size $2\Delta\gtrsim k_{\rm B}T_{\rm c}$, which is a substantial portion of the BCS value. Such flat temperature dependence of $\Delta\lambda$ has been reproduced in three crystals measured in this study, one of which has been measured down to 80\,mK (Fig.\:\ref{lambda}e). The data represent completely temperature-independent behaviour below $T/T_{\rm c}\sim0.05$ within a precision of $\sim1$\,{\AA}  (Fig.\:\ref{lambda}e, inset). The fact that we do not observe any Curie upturn at low temperatures is a strong indication that the point defects are essentially nonmagnetic \cite{Cooper96}. In fact, if we use the reported estimate of the defect density \cite{Beek13}, our precision gives an upper limit of magnetic moment of $\sim0.2\mu_{\rm B}$ per Fe defect, which is much smaller than the moment expected in the magnetic spin states of Fe.

Further increase of the defect density leads to another change of $\Delta\lambda(T)$. The $T$-dependence at the lowest temperatures gets steeper with increasing defect dose, which is an opposite trend to the initial change at low doses. This second stage of the changes clearly indicates that  by irradiation we create the low-energy excitations again inside the formed gap. At the highest doses we measured, we observe the $T^2$ dependence, which is a manifestation of the formation of the Andreev bound states expected for the sign-changing order parameter with impurity scattering. The impurity band corresponding to these bound states must overlap the Fermi level in order to cause this effect.

\begin{figure}[t]
\includegraphics[width=1.0\linewidth]{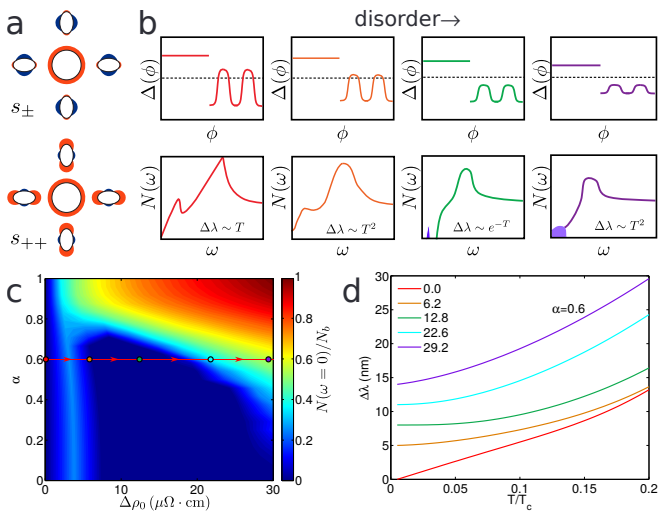}
\caption{{\bf Theoretical calculations of impurity effects in iron-based superconductors.} 
{\bf a}, Schematic of possible $s_\pm$ and $s_{++}$ states. Large circles and small ellipsoids (black lines) are hole and electron Fermi surfaces, respectively. Red and blue represents the superconducting order parameter with different signs. {\bf b},  Schematic of $s_\pm$ order parameter vs. azimuthal angle $\phi$ (top row) and density of states $N$ vs. energy $\omega$ (bottom row) with increasing irradiation dosage (from left to right). Dotted lines (top row) are zero lines.  {\bf c},  Density of states at Fermi level $N(0)$ for nodal band plotted as false color vs. inter/intraband scattering ratio $\alpha$ and irradiation-induced residual resistivity $\Delta \rho_0$ for the model defined in
Ref.\,\onlinecite{Wang13}. {\bf d}, Penetration depth change $\Delta\lambda$ vs. reduced temperature $T/T_{\rm c}$ for different values of $\Delta \rho_0$ as marked in legend, and inter/intraband scattering ratio $\alpha =0.6$ (arrows in {\bf c}). Each curve is shifted vertically for clarity. 
}
\label{theory}
\end{figure}

The observation of the impurity-induced fully gapped state provides one of the strongest pieces of evidence from bulk measurements that the nodes are not symmetry protected in this system. Moreover, the  two-stage changes in the low energy excitations observed here most likely come from the peculiar band-dependent gap functions of multiband iron-pnictide superconductors. Owing to the well separated hole and electron Fermi surface sheets, the fully gapped $s_\pm$ superconductivity with opposite signs of the order parameter on different bands has been predicted by the theories based on spin fluctuations with the antiferromagnetic vector $\bf{Q}=(\pi,\pi)$ in the 2-Fe zone notation \cite{Mazin08,Kuroki09}. The presence of nodes in the pristine samples implies some competing additional mechanisms, such as spin fluctuations with a different $\bf{Q}$ vector \cite{Kuroki09} or orbital fluctuations which prefer the same sign in both bands \cite{Kontani10}. To reproduce the observed two-stage changes, we have made calculations for the low-energy excitations and the low-temperature behaviour of $\lambda(T)$ by using a two-band model \cite{Wang13} (Figs.\:\ref{theory}a-d). We assume that one band is fully gapped, and the other has accidental nodes whose positions are not symmetry protected. Two possible realizations with $s_\pm$ and $s_{++}$ structure are shown in  Fig.\:\ref{theory}a.   Figure\:\ref{theory}b shows schematically the evolution of the gap function in the $s_\pm$ case as a function of increasing disorder of dominant intraband and subdominant interband character.  The corresponding densities of states are also plotted.  Only in the last panels, where the subgap bound state appears,  do the $s_\pm$ and $s_{++}$ cases differ qualitatively.

Figures\:\ref{theory}c and d show concrete theoretical calculations using the model of  Ref.\,\onlinecite{Wang13} in support of this scenario.   The variation of the density of states at the Fermi level with scattering rate and the ratio $\alpha$ of inter- and intraband scattering (Fig.\:\ref{theory}c), clearly shows the nonmonotonic behaviour of the density of states with scattering rate at a fixed $\alpha$.    
The  temperature dependence of $\Delta\lambda$ depends directly on the residual density of states at the Fermi level, and therefore changes with increasing scattering as follows: (1) $T$-linear, (2) $T^2$,  (3) exponential, and (4) $T^2$ (Fig.\:\ref{theory}d). It is the overlap of the bound state in the last panel  -- not present in the $s_{++}$ case -- with the Fermi level that gives rise to the ``re-entrant'' final $T^2$ dependence of the penetration depth. Thus the observed sequential changes of $\Delta\lambda(T)$ in this system are fully consistent with a sign changing $s$-wave superconducting gap of $s_\pm$ type.

Our results demonstrates that the gap topology and the low-energy excitations can be changed by controlling disorder. Such an impurity effect is unprecedented among superconductors, highlighting a unique aspect of iron-based superconductors.  Our study also shows that the impurity effects on gap structure can provide phase information on the superconducting order parameter in bulk measurements, in contrast to other phase sensitive experiments \cite{Tsuei00,Hanaguri10}, most of which require excellent surfaces or interfaces.  

\section*{Methods}
The single crystals of BaFe$_2$(As$_{1-x}$P$_x$)$_2$ were grown by the self-flux method \cite{Kasahara10} and characterised by several techniques as reported previously \cite{Shibauchi14}. The observation of the quantum oscillations in this series of crystals and the sharp superconducting transition indicate the very high quality of our pristine crystals. We used several crystals from three batches, which exhibit slightly different $T_{\rm c0}$ values of 28, 29, and 30\,K. 
Electron irradiation experiments were performed on SIRIUS platform operated by LSI at Ecole Polytechnique, composed by Pelltron type NEC accelerator and closed cycle cryocooler maintaining sample immersed in liquid hydrogen at 20-22\,K during irradiation. The low-temperature environment is important to prevent defect migration and agglomeration. Partial annealing of introduced defects occurs upon warming to room temperature and sample transfer \cite{Prozorov14}.
The resistivity measurements were performed at LSI, Ecole Polytechnique by the van der Pauw method with four contacts on corners of crystals to minimize the possible effect of the unirradiated area due to contacts. The penetration depth measurements by using 13\,MHz TDOs \cite{Hashimoto10,Hashimoto12} were performed at Kyoto University before and after irradiation.

\section*{Acknowledgments}
We thank C.\,J. van der Beek, A. Carrington, H. Kontani, and R. Prozorov for fruitful discussion. We also thank B. Boizot, J. Losco, and V. Metayer for technical assistance. This work was supported by Grants-in-Aid for Scientific Research (KAKENHI) from Japan Society for the Promotion of Science (JSPS), and by the ``Topological Quantum Phenomena'' (No.\,25103713) Grant-in Aid for Scientific Research on Innovative Areas from the Ministry of Education, Culture, Sports, Science and Technology (MEXT) of Japan. Irradiation experiments were supported by EMIR network, proposal No.\,11--10--8071. 

\section*{Author contributions}
T.S. conceived the project. S.Kasahara carried out sample preparation. Y.Mizukami, M.K., S.Kasahara, and T.S. performed irradiation experiments. M.K. performed resistivity measurements. Y.Mizukami, Y.K., S.Kurata, K.H., and T.S. performed penetration depth measurements and analysed data. V.M., A.K., Y.W., and P.J.H. performed theoretical calculations. Y.Matsuda and T.S. directed the research. T.S. wrote the manuscript with inputs from Y.Mizukami, M.K., K.H., V.M., A.K., Y.W., P.J.H., and Y.Matsuda.

\end{document}